\documentclass[10pt,conference]{IEEEtran}

\usepackage[T1]{fontenc}
\usepackage{newtxmath}
\usepackage{amsmath,amssymb}
\usepackage{url}
\usepackage[pdftex]{graphicx}
\usepackage[nobreak]{cite}

\usepackage{framed}
\newcommand{\Commit}[1]{\texttt{#1}}
\newcommand{\tuple}[1]{\langle #1 \rangle}
\def\Range{\mathit{range}}
\def\buildOK{\checkmark_{\!\!\mathit{build}}}
\def\CC{\mathit{CC}}
\def\CN{\mathit{CN}}
\def\All{\mathit{all}}
\def\Changes{\mathit{chg}}
\def\Einit{E_\mathit{init}}

\title{The Impact of Systematic Edits in History Slicing}

\author{%
  \IEEEauthorblockN{Ryosuke Funaki, Shinpei Hayashi, and Motoshi Saeki}
  \IEEEauthorblockA{%
    Tokyo Institute of Technology, Tokyo 152--8550, Japan\\%
    Email: \{rfunaki,hayashi,saeki\}@se.cs.titech.ac.jp%
  }
}

\begin{document}

\maketitle

\begin{abstract}
While extracting a subset of a commit history, specifying the necessary portion is a time-consuming task for developers.
Several commit-based history slicing techniques have been proposed to identify dependencies between commits and to extract a related set of commits using a specific commit as a slicing criterion.
However, the resulting subset of commits become large if commits for systematic edits whose changes do not depend on each other exist.
We empirically investigated the impact of systematic edits on history slicing.
In this study, commits in which systematic edits were detected are split between each file so that unnecessary dependencies between commits are eliminated.
In several histories of open source systems, the size of history slices was reduced by 13.3--57.2\% on average after splitting the commits for systematic edits.
\end{abstract}

\section{Introduction}\label{s:intro}

Version control systems such as Git \cite{git} are widely used for software maintenance.
The use of version control systems enables developers to easily manage product releases and merge the implementation of new features by pull requests.

Developers are sometimes required to extract a subset of their history from the whole \cite{li-tse18}, e.g., while \emph{backporting} changes from one branch to another or applying \emph{branch refactoring} to decompose a branch into multiple subsets. \cite{li-tse18}.
In backporting, major changes such as adding features or fixing bugs are committed to the main branch, and only the necessary subset is cherry-picked and applied to maintenance branches.
In branch refactoring, in situations where unrelated changes are committed together in a single branch, developers untangle the commits in the branch and split them into multiple ones to improve understandability and portability.
This refactoring activity is done while extracting only the necessary changes to create pull requests.

The extraction of a subset of commits may fail \cite{ray-ase13}.
Because changes in commits depend on each other, extracting a subset from a history must take such dependencies into consideration, i.e., if a commit $c$ is included in the subset of changes, another commit $c'$ on which $c$ depends must also be included.
Because the manual identification of these dependencies is a time-consuming task for developers, an automated mechanism is required.

Applications of a slicing to a history structure~(hereafter, \emph{history slicing}) \cite{servant-fse12,servant-icse17,maruyama-ieicet16,li-tse18} are useful in automating the extraction of a near-sufficient set of required changes.
These approaches were inspired by the concept of program slicing \cite{weiser-tse84}, which extracts a set of statements that might affect the value of a variable of interest at a specified program point from the code of a program, to be used as a slicing criterion.
This extracted code is called a \emph{slice} and is computed by a graph reachability algorithm for the program dependence graph of a target program.
In history slicing, a dependence graph of change elements of a history is prepared, and a subset is extracted as a \emph{history slice}.
Change elements of different types, e.g., lines \cite{servant-fse12,servant-icse17}, edit operations \cite{maruyama-ieicet16}, or commits \cite{li-tse18}, can be considered as the application target of the history slicing.

Commits of larger sizes with a wider spread of dependencies can lead to a substantial increase in the size of history slices.
In particular, commits consisting of independent sets of code lead to unnecessary dependencies among changes.
Examples of such changes are non-essential changes \cite{kawrykow-icse11}, tangled changes \cite{herzig-msr13,barnett-icse15,kirinuki-icpc14,sarocha-compsac18}, and impure refactoring changes \cite{gorg-iwpc05,jmatsu-iwpse15}.
In addition, \emph{systematic edits} \cite{kim-icse09,meng-pldi11}, i.e., similar edits to multiple locations of source code, often happen in a history and are troublesome to history slicing.
Because history slices are required to be as small as possible, avoiding unnecessary magnification is important.
However, the impact of systematic edits in history slicing has not yet been investigated.

The purpose of this paper is to investigate the impact of systematic edits on commit-based history slicing.
In particular, we will answer the research question: \emph{How much do systematic-edit-based commits impact the size of commit-based history slices?} (the first contribution).
Based on the empirical study of two open-source systems in Apache Commons ecosystem, we found that a reasonable number of unnecessary changes were included.
To conduct this study, we enhanced commit-based history slicing to make a traditional approach systematic-edits-aware (the second contribution).

The rest of this paper is organized as follows.
Section~\ref{s:example} explains our motivation using a concrete example.
Section~\ref{s:tech} explains our enhancement of commit-based history slicing.
Section~\ref{s:eval} shows an empirical study to answer the research question.
Section~\ref{s:conc} concludes this paper.

\section{Motivation}\label{s:example}

\begin{figure}[tb]\centering
  \includegraphics[height=3.5cm]{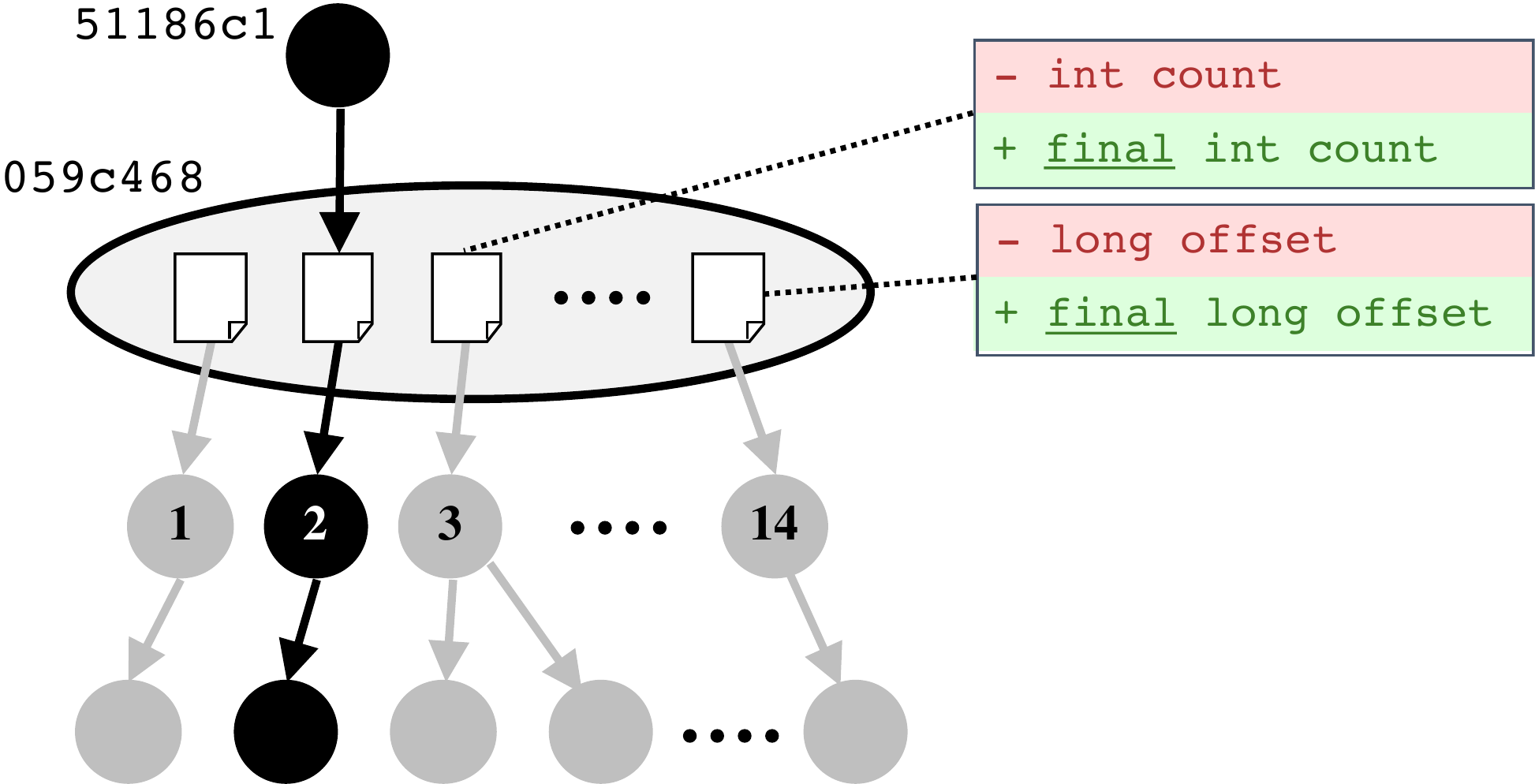}
  \caption{Example of history slicing for Commit \Commit{059c468}.}\label{fig:example}
\end{figure}

Commits consisting of unrelated changes lead to the enlargement of history slices.
Consider the example shown in Fig.~\ref{fig:example}, which illustrates a history slice obtained from Apache Commons Collections.
The commit of \Commit{51186c1}\footnote{\url{https://github.com/apache/commons-collections/commit/51186c1}} is used as the slicing criterion for the commit history of versions 4.1--4.2, which consists of 111 commits.
This commit depends on another commit \Commit{059c468}\footnote{\url{https://github.com/apache/commons-collections/commit/059c468}}, which adds the qualifier \texttt{final} to many fields and variable declarations over 82 files.
Among these, the changed lines in 14 files overwrote other changes in other commits, which implies textual dependencies on 14 changed files.
This means that it is necessary to capture 14 additional commits to capture \Commit{059c468}, which leads to the enlargement of the resulting history slice.
This issue is owing to the commit level granularity of sliced elements in history slicing, i.e., all changes of \Commit{059c468} were included.
In fact, because each addition of the \texttt{final} qualifier does not affect the others, they are independently applicable.
If file changes in the commit can be handled independently on a changed file level, the total number of changes included in the resulting history slice will be reduced, and the resulting slice will become smaller.

\section{Systematic-Edit-Aware History Slicing}\label{s:tech}

\subsection{Overview}

\begin{figure}[tb]\centering
  \includegraphics[width=\linewidth]{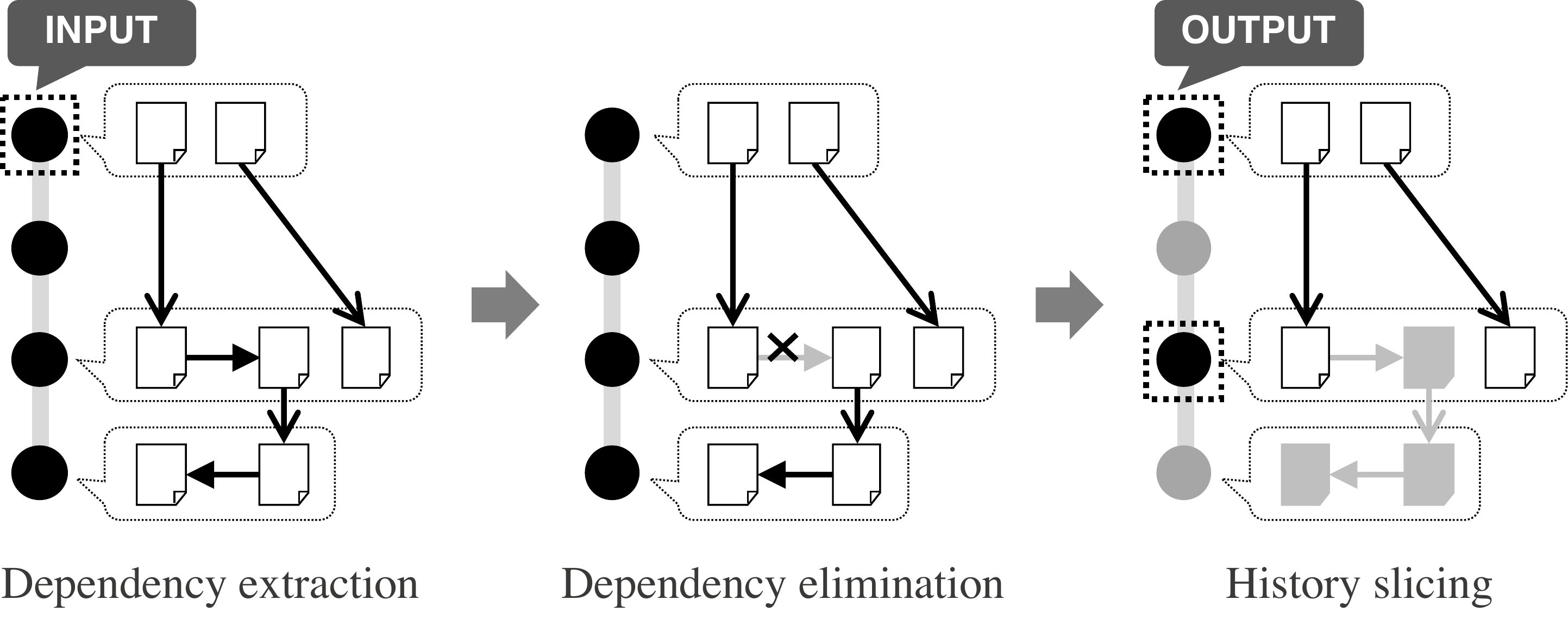}
  \caption{Overview of our history slicing.}\label{fig:overview}
\end{figure}

An overview of our technique of systematic-edit-aware commit-based history slicing is shown in Fig.~\ref{fig:overview}.
Its input is the target commit history and a specific commit for the slicing criteria.
Its output is a history slice consisting of a subset of changed files in the given history.
This process consists of three steps: \emph{dependency extraction}, \emph{dependency elimination}, and \emph{history slicing}.
During dependency extraction, the dependencies between all change elements in the target commit history are extracted.
Dependency elimination detects the commits consisting of systematic edits and eliminates dependencies of a specific type that are regarded as unnecessary.
In history slicing, we apply history slicing using the change elements in the given commit as slicing criteria to obtain a set of related change elements.

\subsection{Dependency Extraction}

In our approach, the basic elements in history slicing are \emph{files} modified in commits rather than \emph{commits} that are used in existing techniques \cite{li-tse18}.
We aim at a finer-grained extraction of history slices because we think that the changes in some commits are inappropriate to be extracted at once.

We call a pair $e = \tuple{c, f} \in E$ of a commit $c \in C$ and a file modified in the commit $f \in F$ \emph{change element}.
The dependencies between change elements are expressed as a binary relation ${\to} \subseteq E \times E$.
We define the dependencies as a union of three different types of dependencies.

\begin{itemize}
  \item A \textit{textual dependency} between change elements $e \to_h e'$ exists if the modified line range, which includes its context lines before and after modification, in $e$ overlaps with that in a past change $e'$.
  More specifically, the dependency can be expressed as
  \[ {\to}_h = \{\, \tuple{e, e'} \mid \Range(e) \cap \Range(e') \neq \emptyset \,\} \]
  where $\Range(e)$ is the modified line range of $e$.

  \item A \textit{build dependency} between change elements $e \to_b e'$ exists if $e'$ are essential for the successful build of the final snapshot of any sub-history including $e$.
  More specifically, the dependency can be expressed as
  \[ {\to}_b = \{\, \tuple{e, e'} \mid \forall E' \subseteq E \bullet e \in E' \land \buildOK(E') \Longrightarrow e' \in E' \,\} \]
  where $\buildOK(E')$ expresses that the build succeeds with the final snapshot of the sub-history $E'$.

  \item A \textit{commit dependency} between change elements $e \to_c e'$ exists if they belongs to the same commit:
  \[ {\to}_c = \{\, \tuple{\tuple{c,f}, \tuple{c',f'}} \mid c = c' \,\}. \]
\end{itemize}

\begin{table*}[tb]\centering
  \caption{Target Systems}\label{tab:system}
  \begin{tabular}{lllllrr} \hline
    Project & History & \hspace{-1.3em}(Versions)   & First commit                    & Last commit     & \# Commits & \# Systematic \\\hline
    Commons Collections
          & $\CC_{4.0}$ & \hspace{-1.3em}(4.0--4.1) & \Commit{5950eba} (2013--11--21) & \Commit{a7cbb44} (2015--11--26) & 211  & 13 \\
          & $\CC_{4.1}$ & \hspace{-1.3em}(4.1--4.2) & \Commit{1ce3b3e} (2015--11--28) & \Commit{483cbbb} (2018--07--08) & 170  & 35 \\
          & $\CC_\All$  & \hspace{-1.3em}(1.0--4.2) & \Commit{3f06f58} (2001--07--15) & \Commit{483cbbb} (2018--07--08) & 2,971 & 330 \\
    Commons Net
          & $\CN_{3.3}$ & \hspace{-1.3em}(3.3--3.4) & \Commit{dc0da97} (2013--06--08) & \Commit{6f97833} (2015--11--19) & 270  & 31 \\
          & $\CN_\All$  & \hspace{-1.3em}(1.0--3.6) & \Commit{564a20c} (2003--02--19) & \Commit{e207b99} (2017--02--11) & 1,954 & 146 \\\hline
  \end{tabular}
\end{table*}

\subsection{Dependency Elimination}

Among the commit dependencies obtained in the previous step, those eliminable are identified.
If there is no semantic relationship between change elements that occurred at the same commit, the commit dependencies between them can be eliminated.

\emph{Systematic edits} \cite{kim-icse09,meng-pldi11} are considered a representative example of changes that do not depend on each other but are committed at once.
Systematic edits are typically done automatically by tools such as those provided by IDEs\@.
The resulting changes in a certain file are considered to not affect the other files of the same systematic editing.
For example, the set of changes adding the qualifier \texttt{final} shown in Fig.~\ref{fig:example} is a typical instance set of systematic edits.

Molderez et al.~\cite{molderez-msr17} proposed a technique to detect systematic edits by abstracting edit scripts of source code changes obtained from differencing abstract syntax trees~(ASTs) of Java source code and grouping them by applying frequent itemset mining.
In this technique, an edit script is expressed as a tuple of $\tuple{\mathit{changeType}, \mathit{structuralSubject}, \mathit{location}}$ where $\mathit{changeType}$ is the type of the change, $\mathit{structuralSubject}$ is an abstracted representation of the AST node to be changed, and $\mathit{location}$ is the absolute path of the node.
The mining method is applied to a set of obtained edit scripts, and the mined frequent patterns are regarded as systematic edits.

We detect systematic edits using Molderez et al.'s technique with an extension.
Our aim is to precisely find \emph{splittable} commits to decrease the dependencies among change elements.
To achieve the accurate detection of such commits, we only focus on commits that can be represented as a single set of systematic edits, i.e., all the changes in the target commit belong to the same instance set of systematic edits.
For this purpose, we do not apply the frequent pattern mining.
We apply an abstraction to edit scripts and check whether all the scripts for all the changed methods follow the same manner.
More specifically, we regard a set of edit scripts at the commit $c$ as systematic if the following condition holds:
\[
  \forall m_1, m_2 \in M_c \bullet \Changes_c(m_1) = \Changes_c(m_2)
\]
where $M_c$ is a set of changed methods at $c$ and $\Changes_c(m)$ is the set of abstracted edit scripts for the changes in $m$ at $c$.

For the abstraction to be applied to edit scripts, we slightly changed the approach of Molderez et al.:
\begin{itemize}
  \item We omit the information on the absolute path stored in \textit{location} because we want to regard changes of different types at different locations to be similar.
  See again the example shown in Fig.~\ref{fig:example}; the addition of the qualifier \texttt{final} occurred at identifiers of different types such as local variables or formal parameters of methods.
  This rule can regard them as the same.
  \item The pair of code fragments before and after the code change is used as the information representing the target AST node stored in \textit{structuralSubject}.
  Molderez et al.'s technique used only either of them: code fragments after change for addition and modification; those before change for deletion.
  This was insufficient to check whether all the changes applied in the same commit satisfied the above conditions.
\end{itemize}
We regarded the changes of a commit as systematic if all the abstracted edit scripts of them follow the conditions shown above.
In addition, commits consisting of only white-spaces or comment changes without any syntax tree differences are also systematic of a special type.

If commits consisting of systematic edits are found, we collect the commit dependencies of these commits $\tilde{\to}_c$.
These commit dependencies are eliminated from the set of all dependencies: ${\to'_c} = {\to_c} \setminus \tilde{\to}_c$.
The history slices are then calculated using the updated dependencies.

\subsection{History Slicing}

For the change elements contained in the input commits $\Einit$, this step recursively traces the dependencies obtained in the previous steps ${\to} = {\to_h} \cup {\to_b} \cup {\to'_c}$ and retrieves all the change elements that are traceable by the dependencies as a history slice:
\[
  E^* = \bigcup_{e \in \Einit} \{\, e' \mid e \to^* e' \,\}
\]
where ${\to}^*$ is a transitive closure of the dependencies ${\to}$.

The list of commits is then extracted based on the obtained slice.
All the non-systematic-edit-based commits containing at least one change element in $E^*$ are cherry-picked and included in the output.
For systematic-edit-based commits, they are reconfigured so only the necessary change elements are included and are committed to the output.

\section{Preliminary Study}\label{s:eval}

\subsection{Implementing the Approach}\label{s:impl}

We implemented a history slicing tool to realize the proposed technique.
The technique targets Java projects managed by Git and handles dependencies on Java source code.
We used JGit \cite{jgit} to manipulate the change histories obtained from Git repositories.
For extracting textual dependencies, we used CSlicer \cite{li-tse18}, which used the git-deps algorithm \cite{gitdeps}.
For extracting build dependencies, we used GumTree \cite{falleri-ase14} to extract fine-grained changes from commits and find def-use relations in the obtained changes.
We used Eclipse Java Development Tools \cite{jdt} to search for the definition and use of identifiers.

\subsection{Data Collection}

To answer the stated research question, we applied the implemented history slicing to several histories of open-source systems written in Java from the Apache Commons ecosystem.
Table~\ref{tab:system} shows the histories and systems we selected.
The histories were selected as in-between two releases of the projects.
The ends of this interval are termed first and last commits, as indicated in the table.
For the simplicity of the input histories, we followed the selection criterion that there were no merge commits during the interval between two releases.

\subsection{Data Analysis}

We applied our proposed history slicing for each commit in the given history, i.e., each commit is regarded as the slicing criterion, and a history slice is obtained according to the slicing criterion.
For example, because the history $\CC_{4.0}$ consists of 211 commits, we obtained 211 slices, and each of them was a subset of the given original history.

\subsection{Results and Discussions}

\begin{table}[tb]\centering
  \caption{Accuracy in Detecting Systematic Edits}\label{tab:sysed}
  \begin{tabular}{lrrrrr} \hline
    History     & \# Detected & \# Correct & \# Independent \\ \hline
    $\CC_\All$  & 330 ($82+248$) & 78 ($78/82 = 95$\%) & 73 ($73/78 = 96$\%) \\
    $\CN_\All$  & 146 ($32+114$) & 31 ($31/32 = 97$\%) & 29 ($29/31 = 94$\%) \\ \hline
  \end{tabular}
\end{table}

As a preliminary study, we investigated the accuracy of our systematic edits detection.
The result is shown in Table~\ref{tab:sysed}.
We applied all the commits in Commons Collections~($\CC_\All$) and Commons Net~($\CN_\All$).
Among the 2,971 and 1,954 commits in $\CC_\All$ and $\CN_\All$, 330 and 146 were detected as systematic edits, respectively.
Note that 82 of 330 and 32 of 146 commits included AST differences; the others only included white-space and comment changes.
One of the authors manually confirmed whether these AST changes are actually systematic edits and if the edits in them are actually independent of each other.
As a result, 78 and 31 of the detected systematic edits were correct, which suggests a detection precision of 95\% and 97\%, respectively.
Furthermore, 73 and 29 of the correct systematic edits were regarded as splittable, i.e., each edit in them was independent of each other.
This result suggests the validity of using systematic edits detected by our approach to split commits.
Note that for two of the commits that are regarded as unsplittable, the necessary dependencies were covered by the build dependencies; thus, the resulting history slices were correct even though these non-splittable commits were regarded as systematic edits.

\def\FigHeight{3.95cm}
\begin{figure}[tb]\centering{\footnotesize
  \includegraphics[height=\FigHeight]{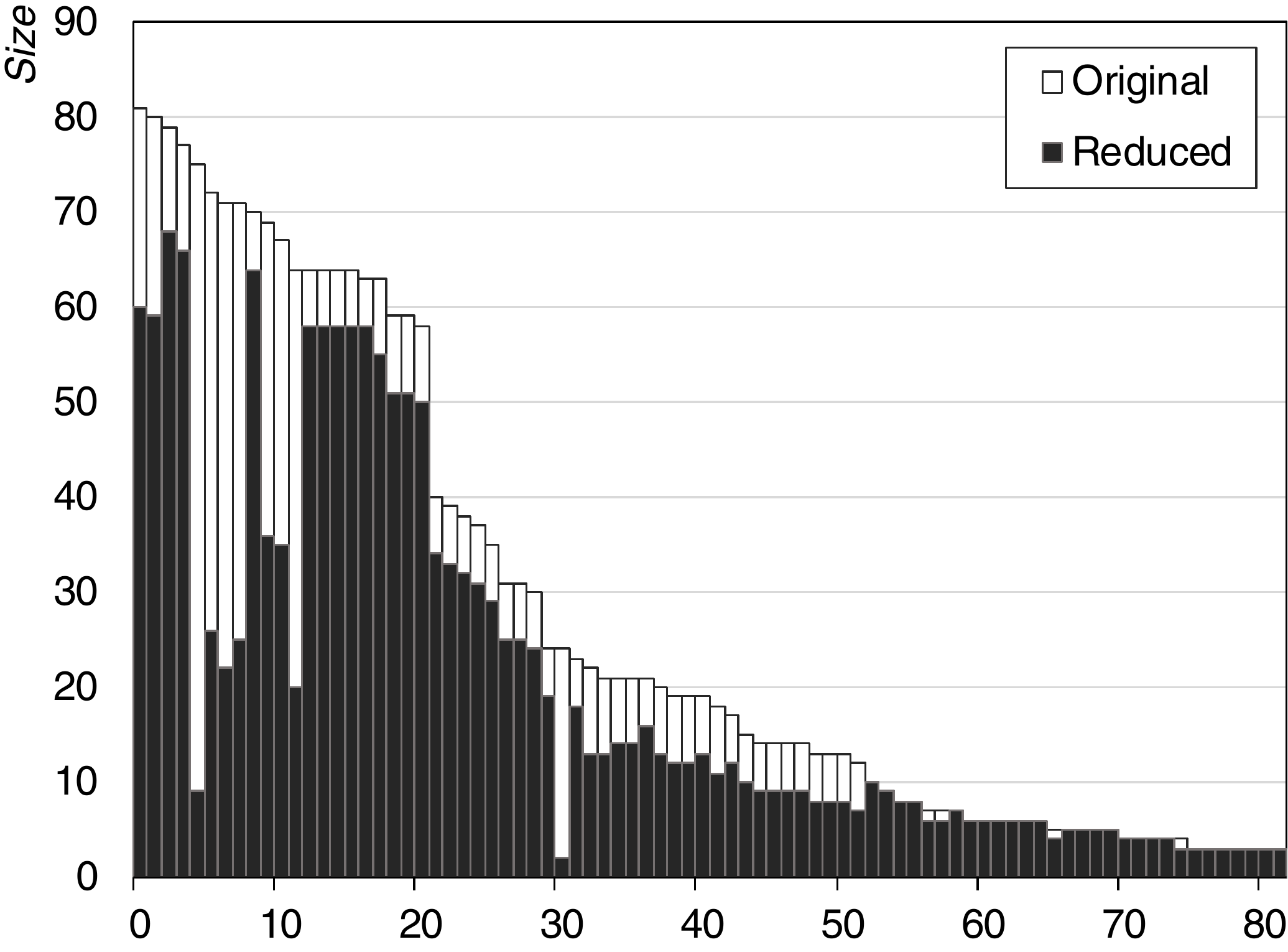}\hspace{0.5cm}
  \includegraphics[height=\FigHeight]{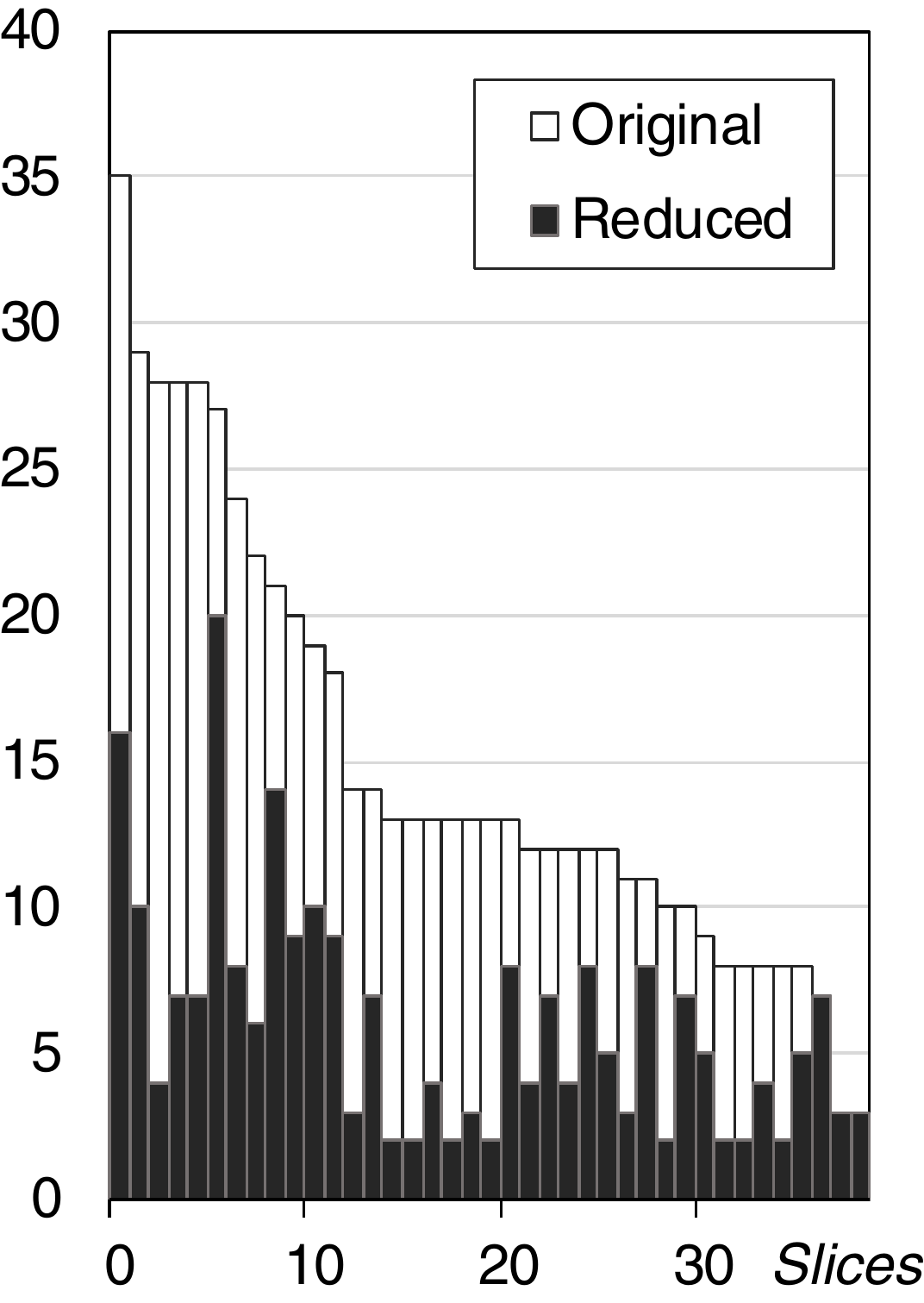}\\
  \hspace{0.15cm}
  (a) $\CC_{4.0}$: Commons Collections 4.0--4.1 \hspace{0.7cm}
  (b) $\CC_{4.1}$: 4.1--4.2 \vspace{1em}\\
  \includegraphics[height=\FigHeight]{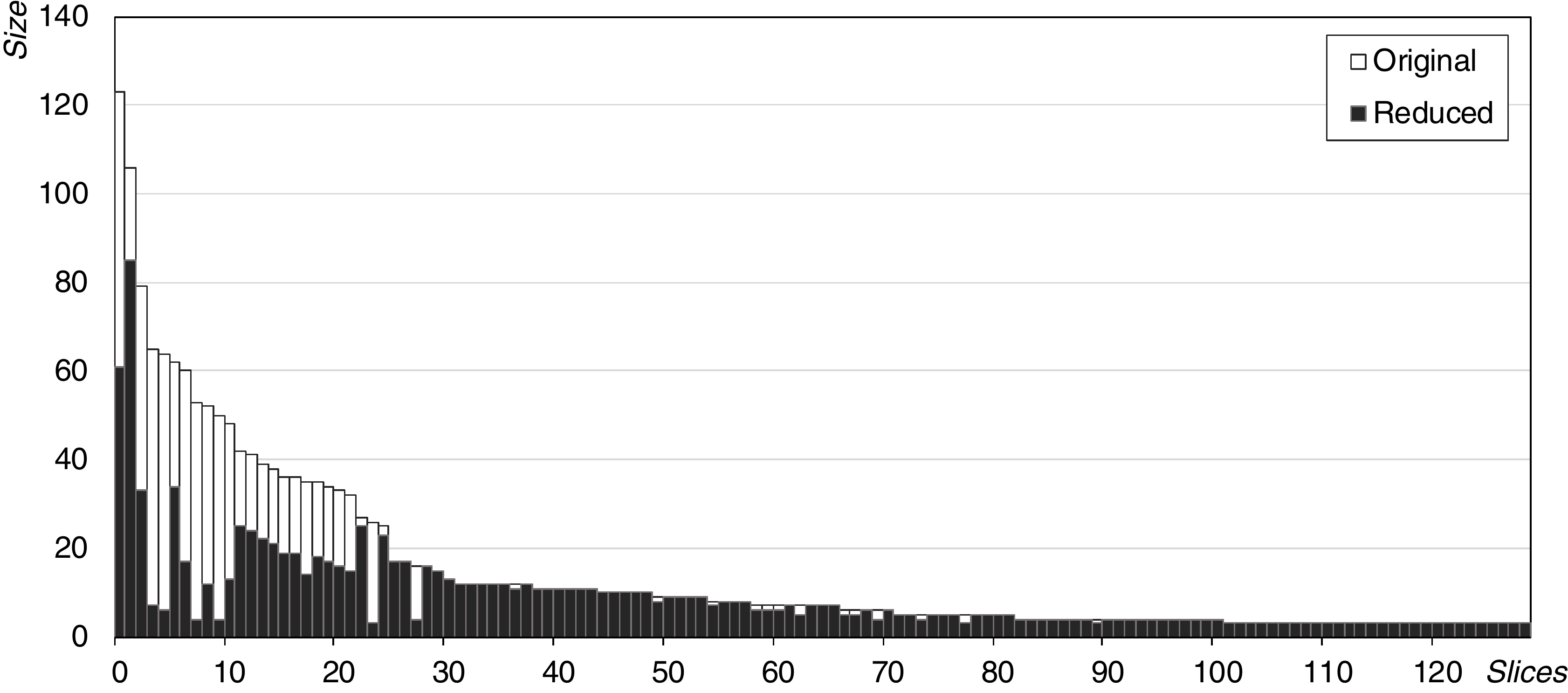}\\%
  (c) $\CN_{3.3}$: Commons Net 3.3--3.4.%
  }
  \caption{Impact of systematic edits on history slices.}\label{fig:result}
\end{figure}

The impact of systematic edits on the size of history slices for each history are shown in Fig.~\ref{fig:result}.
For each bar-plot, the bar shows a history slice, whose vertical size represents its size.
All the slices are shown horizontally and are sorted in descending order by their size in the vertical.
Slices whose sizes were less than three were excluded because they were too small to consider while discussing the reduction impact.
The ``Original'' bars represent the slice size without applying dependency elimination with systematic edits detection, whereas the ``Reduced'' bars represent the size with the application of dependency elimination.
Further, the numbers of detected sets of systematic edits are shown in the ``\# Systematic'' column in Table~\ref{tab:system}.

A reduction rate was observed in the size of the slices of $\CC_{4.0}$, $\CC_{4.1}$, and $\CN_{3.3}$ by 20.7\%, 57.2\%, and 13.3\% in average, respectively.
The figure also shows that reduction succeeded at most slices but not for specific small subsets of them.
This result suggests the importance of the consideration of systematic edits in the size reduction of history slices.

A history slice of $\CN_{3.3}$ whose size was 123 obtained using the commit \Commit{ab2fd4f} as the slicing criterion was reduced to one whose size was 61 after enabling systematic edits detection.
We confirmed the extent of change elements with widely spread dependence for this slice.
We found that there were seven commits in the slice that textually depended on more than 10 commits.
The severest commit was \Commit{4140189}, which had textual dependencies to 26 other commits.
The elimination of dependencies did not allow the slicing criterion to reach these severe commits, which led to the size reduction of the resulting slice.

\begin{framed}
  \noindent
  The size of history slices decreased by 13.3--57.2\% on average when splitting commits consisting of systematic edits.
\end{framed}

\section{Conclusion}\label{s:conc}

This paper proposes an enhancement for history slicing based on the detection of systematic edits.
By extracting the dependency relationships among change elements and eliminating some of them if they are detected as systematic edits, developers are able to obtain history slices of a reduced size.
The conducted empirical studies show that the elimination of commit dependencies in the commits related to systematic edits reduces the size of resulting history slices by 13.3--57.2\% on average.

As future work, several threats to the validity of the conducted empirical studies, e.g., generality of the selected projects and histories for the external validity, and the accuracy of the extracted dependencies for the internal validity, should be mitigated.
Furthermore, we will further investigate the impact of commits consisting of independent sets of code by looking at the other types of changes.

\section*{Acknowledgments}
This work was partly supported by Kayamori Foundation of Informational Science Advancement and JSPS KAKENHI Grant Numbers JP18K11238, JP15K15970, JP15H02683, and JP15H02685.

\IEEEtriggeratref{12}


\end{document}